\newcommand{\CLASSINPUToutersidemargin}{19mm}
\newcommand{\CLASSINPUTtoptextmargin}{19mm}
\newcommand{\CLASSINPUTbottomtextmargin}{19mm}

\documentclass[10pt,conference]{IEEEtran}

\IEEEoverridecommandlockouts

\usepackage[noadjust]{cite}
\usepackage{amsmath,amssymb,amsfonts}
\usepackage{algorithmic}
\usepackage{graphicx}
\usepackage{textcomp}
\usepackage{xcolor}

\usepackage{amsthm}
\usepackage{mathabx}

\newtheorem{theorem}{Theorem}[section]

\newtheorem{lemma}[theorem]{Lemma}

\newtheorem{corollary}[theorem]{Corollary}

\newtheorem{proposition}[theorem]{Proposition}

\newcounter{chap}
\newcounter{sect}

\def\NN{\mathbb{N}}

\def\RR{\mathbb{R}}

\begin{document}

\title{The Option Pricing Model Based on Time Values:\\An Application of the Universal Approximation Theory on Unbounded Domains}

\author{\IEEEauthorblockN{Yang Qu\textsuperscript{\textsection}}
\IEEEauthorblockA{\textit{School of Mathematics, Hunan University} \\
Changsha, China \\
quyang@hnu.edu.cn}
\and
\IEEEauthorblockN{Ming-Xi Wang\textsuperscript{\textsection}}
\IEEEauthorblockA{\textit{PAG Investment Solutions}\\
Geneve, Switzerland \\
mingxi.waeng@gmail.com}
}

\maketitle
\begingroup\renewcommand\thefootnote{\textsection}
\footnotetext{Equal contribution}
\endgroup

\begin{abstract}
We propose a time value related decision function to treat a classical option pricing problem raised by Hutchinson-Lo-Poggio. In numerical experiments, the new decision function significantly improves the original model of Hutchinson-Lo-Poggio with faster convergence and better generalization performance. By proving a novel universal approximation theorem, we show that our decision function rather than Hutchinson-Lo-Poggio's can be approximated on the entire domain of definition by neural networks. Thus the experimental results are partially explained by the representation properties of networks.
\end{abstract}

\section{Introduction}
Option pricing is an important topic in financial mathematics. Black and Scholes \cite{BS} and Merton \cite{Merton} obtained celebrated valuation formulas of closed-form under the assumption of some ideal conditions. However, these conditions hardly hold in reality. Thus it is necessary to investigate alternative pricing methods in actual trading conditions. Hutchinson, Lo and Poggio \cite{HutchinsonLoPoggio} proposed approaching the valuation of options with neural networks and compared their model with the Black-Scholes formula through numerical experiments. This pioneering work has motivated many further investigations along this line \cite{ruf2019neural}.

It is common to evaluate an option pricing model by comparing it with the Black-Scholes formula. Hutchinson et al. raised the following question in \cite{HutchinsonLoPoggio}: ``If option prices were truly determined by the Black-Scholes formula exactly, can learning networks learn the Black-Scholes formula?'' At time $t$, the price $C(t)$ of a European call option with maturity $T$ is theoretically determined by its underlying asset's price $S(t)$, strike price $K$, time to maturity $\tau = T - t$, dividend yield $q$, risk-free rate of interest $r$, and volatility $\sigma$ (\cite{BS}, \cite{Merton}). Assuming that $q$, $r$, and $\sigma$ are constant through time, the following learning objective

\vspace{-1\baselineskip}
\begin{small}
\begin{align}\label{model1}
(S(t)/K,\tau) \to C(t)/K
\end{align}
\end{small}%
is proposed by Hutchinson et al. in \cite{HutchinsonLoPoggio}. Let $f$ be defined by $f(S(t)/K, \tau) = C(t)/K$. As options actively traded in US exchanges are non-LEAPS ones that are of time to maturity less than one year, without loss of generality we can assume that $f$ is a function defined on $\RR^{+} \times [0,1]$. The fact that $f$ is unbounded on $\RR^{+} \times [0,1]$ suggests the following: the neural network (\ref{model1}) with a bounded activation function (such as the logistic function used in \cite{HutchinsonLoPoggio}) is unlikely to generalize well on calls deep in-the-money, where $S(t)$ is significantly higher than $K$ and thus $f$ becomes exceedingly large. A model without taking this into account is of limited practicality because of its unpredictability during tail events, such as the Volkswagen's spike in October 2008, the jump of the Swiss franc in January 2015, and the global stock market crash caused by COVID-19 in February 2020. Even within the compass of bounded training data, the authors of \cite{HutchinsonLoPoggio} have pointed out that the largest errors of model (\ref{model1}) tend to occur at points corresponding to near-the-money options at expiration and points along the boundary of the sample; and the authors of \cite{YangZhengHospedales} have summarized that model (\ref{model1}) and its variations tend to overestimate deep out of the money options or underestimate options very close to maturity.

To alleviate these issues, we propose an alternative decision function to settle the problem of Hutchinson-Lo-Poggio:

\vspace{-1\baselineskip}
\begin{small}
\begin{align}\label{model2}
(S(t)/K,\tau) \to V(t)/K
\end{align}
\end{small}%
where $V(t)$ is the time value of the call option at time $t$. This paper treats only European options, and thus the time value of a European call option is set to be

\vspace{-1\baselineskip}
\begin{small}
\begin{align}\label{modeleq1}
V(t) = C(t) - (S(t) e^{-q \tau}-Ke^{-r \tau})^+
\end{align}
\end{small}%
where $x^+ := \max{(0,x)}$. Such a decision function is motivated by the perspective of market practitioners: in many cases, the time value provides more valuable information than the option price. In this paper we present a detailed proof of that a shallow artificial neural network with logistic activation is a universal approximator of $L^{2}(\RR \times [0,1])$. Let $g$ be defined by $g(S(t)/K,\tau)= V(t)/K$. We will prove that $g$ is a function in $L^2(\RR^{+}\times [0,1])$. Moreover, $g$ rather than $f$ can be approximated by superpositions of logistic function in $L^2(\RR^+ \times [0,1])$. This result provides a theoretical basis for the better generalization performance of Model (\ref{model2}) in numerical experiments.

\begin{figure}[!t]
\centering
\includegraphics[width=3.5in]{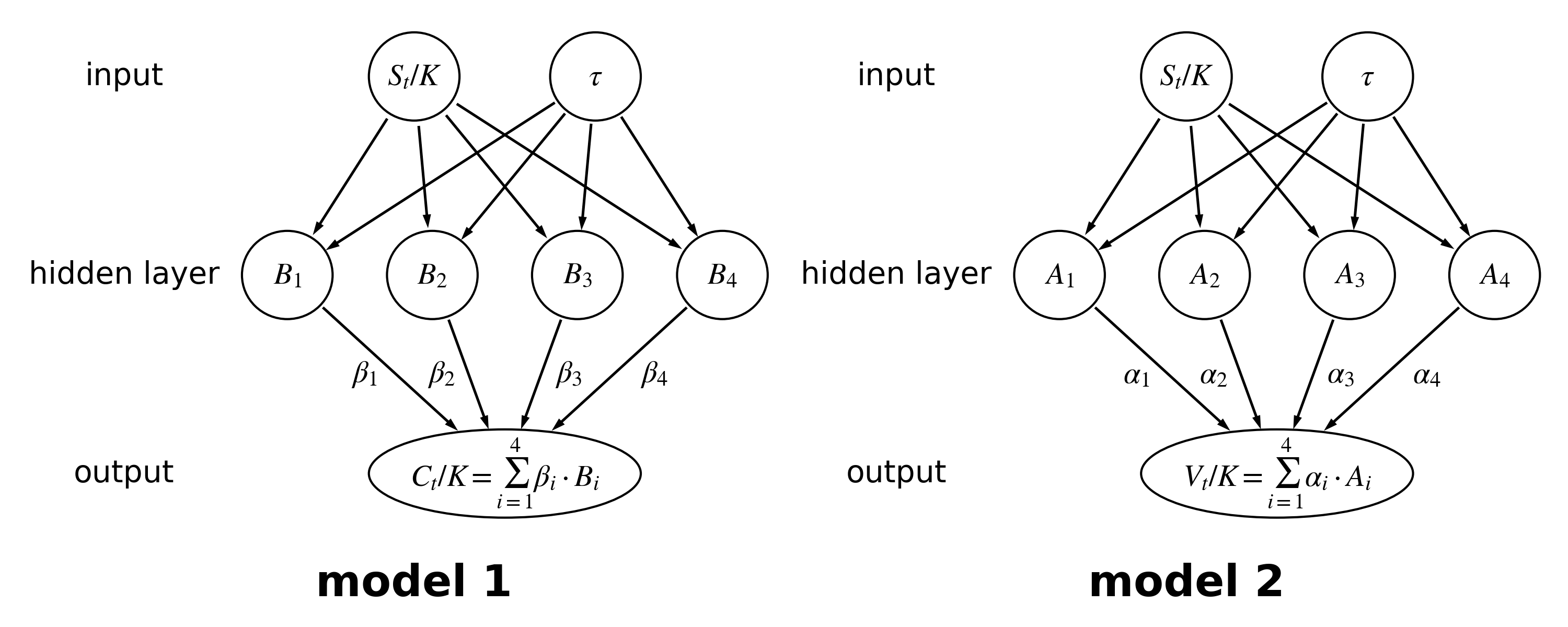}
\caption{The structure of model (\ref{model1}) and model (\ref{model2}). They both have only one hidden layer, and in our numerical experiments we both take four hidden units. The difference between two models is the output variables.}
\label{model_structure}
\vspace{-0.5cm}
\end{figure}

The neural network discussed in this paper is a multilayer perceptron (MLP) with a single hidden layer. Fig. \ref{model_structure} compares the architecture of our model with that of Hutchinson-Lo-Poggio's. Our network attempts to fit the proposed decision function by

\vspace{-1\baselineskip}
\begin{small}
\begin{align*}
g(S(t)/K,\tau) = \sum\limits_{i=1}^{k}\alpha_i h(a_iS(t)/K + b_i\tau - \theta_i)
\end{align*}
\end{small}%
where $a_i$, $b_i$ and $\alpha_i$ are the connection weights, $\theta_i$ the thresholds, $h$ the activation function, and $k$ the number of hidden units. The experimental results demonstrate that Model (2) is faster in convergence and better in out-of-sample performance.

The organization of this paper is as follows. In section \ref{related work}, we discuss a selection of works that are most relevant to this paper. In section \ref{sectiondata}, model (\ref{model1}) and model (\ref{model2}) will be trained and compared on simulated data as well as market data. In section \ref{sectionbs}, we study the boundary behavior of $g$ and show that it is a function in $L^p(\RR^+\times [0,1])$ for all $1 \leq p \leq +\infty.$ In section \ref{sectionnn}, we prove Theorem \ref{maintheorem} which indicates that our model is of better representation power and explains the better generalization performance of the model.

%In section \ref{resnResNetet}, we point out that our model (\ref{model2}) is in a residual fashion similar to that of ResNet \cite{he2016deep} and it can explain why model (\ref{model2}) achieves faster learning. In section \ref{sectionbs}, we study boundary behavior of $g$ and show that it is an element in the $L^p(\RR^+\times [0,1])$ for all $1 \leq p \leq +\infty.$ In section \ref{sectionnn}, we prove Theorem \ref{maintheorem} which implies that our model is better at expressivity and justifies that our model has a better generalization performance.

\begin{table}[!t]
\renewcommand{\arraystretch}{1.1}
\caption{Notations}
\label{table_example}
\centering
\begin{tabular}{ll}
%\hline \bfseries
$t$ & Current time\\
$T$ & Maturity \\
$\tau=T-t$ & Time to maturity\\
$C(t)$ & Price of the call option at $t$ \\
$V(t)$ & Time value of the call option at $t$, see (\ref{modeleq1}) \\
$S(t)$ & Price of the underlying security at $t$ \\
$R(t)$ & $\log(S(t)/S(t-1))$\\
$K$ & Strike price \\
$q$ & Dividend yield\\
$r$ & Risk-free rate\\
$s, d_1,d_2, N(x)$ & see Section \ref{sectionbs} \\
$\sigma$ & Volatility\\
$f, g$ & Functions of two variables, see (\ref{deff}) and (\ref{defg}) \\
$IV$ &  $f-g$ \\
$W(t)$ &  Wiener process \\
$e$ & The base of the natural logarithm\\
$\exp, \log$ & Exponential, natural logarithmic function\\
$I_{U}$ & Indicator function of $U$ \\
$N(\mu, \sigma^2)$ & Normal distribution\\
$x^+$ & $\max{(0,x)}$ \\
$L^p(U)$ &  Space of $L^p$-integrable functions on $U$ \\
$\overline{\mathcal{X}}, \mathcal{X}\subset L^2(U)$ &  Closure of $\mathcal{X}$ in $L^2(U)$ \\
$C(U)$ &  Space of continuous functions on $U$\\
$C_0(\RR^n)$ &  Space of continuous functions on $\RR^n$ vanishing at $\infty$ \\
$\RR$ & Real numbers\\
$\RR^n$ & n-dimensional Euclidean space \\
$\RR^+$ & Positive real numbers\\
$\RR^-$ & Negative real numbers\\
$C_1, \ldots, C_7$ & Constants in estimates \\
$C_{\eta}, C_{\eta}'$ & Constants in estimates \\
$B, X, C$ & Constants in estimates \\
$\langle y, x\rangle, x,y \in \RR^n$ & Dot product of $x$ and $y$ \\
$\psi$ & Some function on $\RR$ \\
$\psi^{\tau_{\theta}}, \psi^{\delta_{y}}, \psi^{\tau_{\theta}\delta_{y}}$ & see (\ref{actions}) \\
$\mathcal{S}_n(\psi)$ & Functions represented by a shallow network, see (\ref{Sn}) \\
$\phi$ & Logistic function \\
$S^1$ & Unit circle $\{x\in\RR^2: \langle x,x \rangle=1 \}$ \\
$\lambda_n$ & Lebesgue measure on $\RR^n$ \\
$\lambda_{S^1}$ & The canonical measure on $S^1$ with volume $2\pi$ \\
$\theta^{\perp}$ where $\theta \in S^1$ & If $\theta = (\cos{\omega}, \sin{\omega})$ then $\theta^{\perp}=(-\sin{\omega}, \cos{\omega})$ \\
$R, S$ & Functions, see (\ref{defRS}) \\
$F \mapsto \widehat{F}$ & Fourier transform \\
$P_{\theta}F$ & X-ray transform of $F$, see (\ref{defxray}) \\
$\Omega$ & $\RR \times [0, 1]$ \\
$\Omega_k$ & $(-k, k) \times [0, 1]$ \\
$h_k$  & $I_{\Omega_k}h$ \\
$||F||_{L^2(U)}$ & The norm of $F$ in $L^2(U)$ \\
$\gamma_{y, \rho}$ & Function, see (\ref{defgamma}) \\
$\theta_{\xi}, B_{\eta}, \Theta_{\eta}$ & see the paragraph before (\ref{defalot})  and Fig. \ref{proofpng} \\
$T^{\pm}_{\theta, k}, T^{\dag}_{\theta, k}, S_{\theta, k, t}$ & see (\ref{defalot}) and Fig. \ref{proofpng} \\
$Y^{\pm}_{\theta, k}, Y^{\dag}_{\theta, k}$ & see (\ref{defalot}) and Fig. \ref{proofpng} \\
$O(s)$ & Bachmann Landau notation in terms of $s$ \\
%\hline
\end{tabular}
\end{table}

\section{Related work}\label{related work}
In this section, we review some relevant works on option pricing networks and universal approximation theorems. In literature, most papers concerning option pricing networks have paid attention to input features but not the decision functions. A recent survey \cite{ruf2019neural} summarized the option pricing networks out of more than 150 papers, among them decision functions based on the option price such as $C(t)/K$ and $C(t)$ are the most popular ones. However, a few authors have tried different decision functions. For example, Boek et al. \cite{Lajbcygier} considered $C(t) - C_{BS}$, which is the deviation of the market price from Black-Scholes price. This deviation is somewhat related to the principle of our paper. Gradojevic et al. \cite{Gradojevic} proposed a divide-and-conquer strategy to improve the performance of option pricing networks on points around the boundary. This strategy has been further developed by Yang et al. \cite{YangZhengHospedales}. Also, the decision function used in \cite{YangZhengHospedales} is $C(t)/S(t)$ instead of $C(t)/K$.

The universal approximation theorem in the mathematical theory of artificial neural networks was established by Cybenko \cite{Cybenko}, with alternative versions and proofs contributed by Hornik-Stinchcombe-White \cite{Hornik1989} and Funahashi \cite{Funahashi}. This classical theory treats only functions on bounded domains, which seems an obstacle for many applications in finance. For it is necessary to take heed of the tail risk there. There are a few authors who have studied the approximation of neural networks on unbounded domains. For example, \cite{Ito} and \cite{Chen1991} have studied the approximation capabilities of neural networks in $C_0(\RR^n)$, which is the space of continuous functions of $\RR^n$ vanishing at the infinity. Hornik \cite{Hornik1991} has studied the approximation capabilities of neural networks in $L^p(\RR^n, \mu)$ concerning a finite input space environment measure $\mu$.  It is of particular interest to investigate the $L^2$ approximation of neural networks, as it is related to the conventional mean squared error. In Theorem \ref{maintheorem}, we prove that a shallow network with logistic activation is a universal approximator in $L^{2}(\RR \times [0,1])$. On a bounded domain $\Omega$, a universal approximation theorem in $C(\Omega)$ already implies a universal approximation theorem in $L^p(\Omega)(p \in (1,\infty))$, as $C(\Omega)$ is a dense subset of $L^p(\Omega)$. However, on an unbounded domain $\Omega$, there is no such simple relationship between $C(\Omega)$ and $L^p(\Omega)$. Therefore, our Theorem \ref{maintheorem} is a complement rather than a corollary of the work of \cite{Chen1991}. Lastly, Theorem \ref{maintheorem} has been generalized to higher dimensions for almost all other popular activation functions in \cite{MWunboundeddomain}.

\section{Numerical experiments}\label{sectiondata}

%We test them using simulated data and real data.

In this section, we evaluate the performance of Hutchinson-Lo-Poggio's Model (\ref{model1}) and our Model (\ref{model2}) in simple settings. For the simulated data, the underlying asset's price is generated according to the Black-Scholes conditions, and the option price is calculated using the Black-Scholes formula. Testing such data can let us know how well the model learns the Black-Scholes formula. For real data that come from the market, testing them allows us to measure the practicality of the model.

In the computational programs we use the framework of tensorflow.keras. We compare them under the same hidden unit number, activation function, optimizer, and iteration number. The optimizer used is Adam \cite{Adam}, as it is well known to experts that Adam is much more forgiving to hyperparameters. We choose mean squared error (MSE) as the loss function because it is the most popular one and it is compatible with our Theorem \ref{maintheorem}. Model (\ref{model1}) in \cite{HutchinsonLoPoggio} was trained for learning networks with four hidden units, and we take four hidden units in our experiments as well.

\subsection{Simulated data}

In this section, we use the method in \cite{HutchinsonLoPoggio} to generate simulated data to test the models. We assume that the underlying asset's price of the simulation experiment satisfies the Black-Scholes assumption, that is, the price is a geometric Brownian motion:

\vspace{-1\baselineskip}
\begin{small}
\begin{align*}
dS(t) = \mu S(t)dt + \sigma S(t)dW(t).
\end{align*}
\end{small}%
We set the initial price $S(0)$ to be 100, the annual continuously compounded expected rate of return $\mu$ to be 10\%, and the annual volatility $\sigma$ to be 20\%. For simplicity, we assume that there are 30 trading days per month and the simulated data lasts for one year. We generate 360 days of daily logarithmic returns $R(t)$ based on the normal distribution $N(\mu/360,\sigma^2/360)$, and the asset's price of date $t$ is:

\vspace{-1\baselineskip}
\begin{small}
\begin{align*}
S(t) = S(0)\cdot \exp(\sum\limits_{i=1}^{t} R(i)).
\end{align*}
\end{small}%

Every day we will be based on the asset's price to decide whether to issue new options. The strike price $K$ of a new option satisfies $0.8 < S/K < 1.25$, and $K$ is a multiple of 5. Each strike price will generate eight different expiration dates within one year: six continuous recent months and the next two quarterly months, and the expiration date is set to be the end of the month. We assume that the existing options will be traded every day until maturity.

The data structure used for simulation contains a total of four variables, including the same two input variables $S/K$ and $\tau$ of model (\ref{model1}) and model (\ref{model2}), the output variable $C/K$ of model (\ref{model1}), and the output variable $V/K$ of model (\ref{model2}). Here we take risk-free rate of interest $r = 0.02$ and dividend yield $q = 0$ in expression (\ref{modeleq1}), and the option price $C$ will be obtained by the Black-Scholes formula. We generate one year of data as a training set across the above method and generate another one year of data as a validation set across the same method. To illustrate the advantages of our model, we use two other types of data sets as the test set:

\begin{itemize}
  \item The first type of test data set consists of options with large or small $S/K$: For this type of data, the underlying asset's price is generated as described above, but the initial price $S(0)$ is far from the strike price, i.e., $S(0)/K$ is very large or small. This kind of option has a special meaning in actual transactions: When there is a large change in the market, such as breaking news or bubble broken led to asset price inflation or collapse, then the $S/K$ of the existing option will become far away from general situations (0.8 -- 1.25). These special cases in the market are rare, so the amount of data available for learning these cases will be a little or zero. This is where we look at our model's learning abilities. In this type of test data set of our experiments, $S(0)/K$ are taken with 0.5 and 2.0, and the data length of an option is one year. In order to obtain more meaningful results, we generate 10 different asset price series from the initial price independently and merge them.

  \item The second type of test data set consists of expiration date's data: We generate a large amount of $S/K$ uniformly distributed between 0.8 and 1.25, and take $t = T$, formed this type of data set. Our aim is to observe the performance of the model on the time border. Here we take the number of data equal to 50000.
\end{itemize}

Therefore, our simulation data contains three sets: training set, validation set and test set. The data of the validation set and the training set have the same distribution, while the test set and the training set have different distributions. The training set is used to iterate the model parameters, and the validation set is used to observe where over-fitting occurs during the iteration. We do not give rules for the termination of iterations, because the goal of this article is not to optimize model calculations. The result of the test set can really tell us whether the model is good or bad.

We use MSE as a measure of the quality of the model. In order to obtain the result of a statistical average, we complete 1000 independent training-testing process, and average the results of all. We calculate 11 different activation functions with Adam optimizer and take batch size 128.

\begin{figure}[!t]
\centering
\includegraphics[width=3.5in]{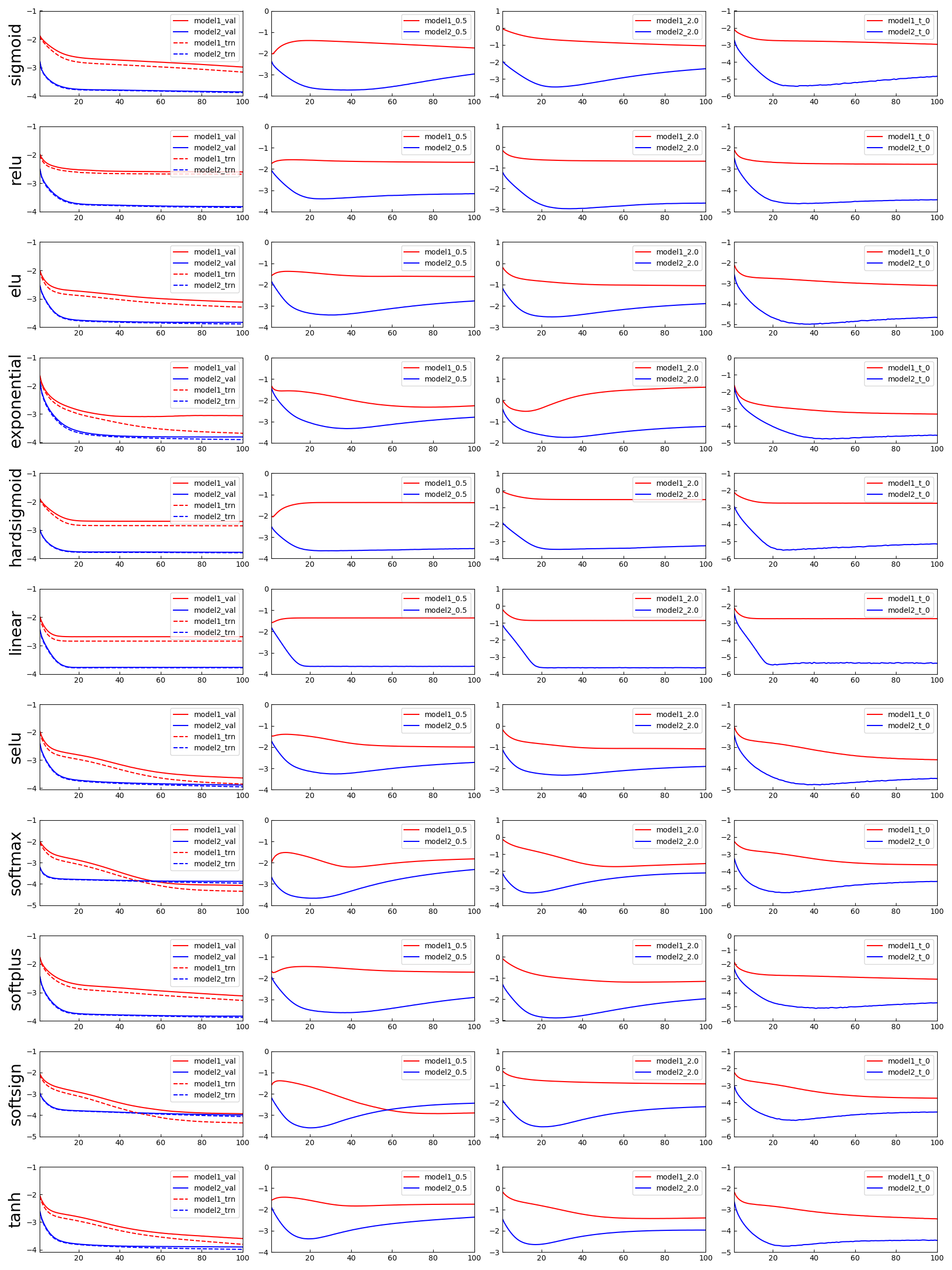}
\caption{The results of simulated data. Each row is the result of the same activation function, with the function name written on the left. The red lines represent the results of model (\ref{model1}) and the blue lines represent the results of model (\ref{model2}). The dashed lines represent the training set, and the solid lines represent the validation set and the test set. The x-axis value of each subfigure is epoch, and the y-axis value is $\log_{10}(\mathrm{MSE})$.}
\label{simresult}
\vspace{-0.5cm}
\end{figure}

Fig. \ref{simresult} shows the MSE with the increase of the epoch in our simulation experiments. It consists of $11 \times 4$ subfigures. The x-axis value of each subfigure is epoch, and the y-axis value is $\log_{10} (\mathrm{MSE})$. We used a total of 11 different activation functions for calculation, and the four subfigures of each row belong to the same activation function. The first column of the figure shows the results of the training set and the validation set; the second and third columns show the results of the test set with $S(0)/K = 0.5$ and $S(0)/K = 2.0$; the fourth column shows the result of the test set consisting of expiration date's data. The red represents the results of Model (\ref{model1}), and the blue represents the results of Model (\ref{model2}). The dashed lines represent the training set, and the solid lines represent the validation set and the test set. It can be seen from the figure that the error of our model is less than the error of Hutchinson-Lo-Poggio's model at each epoch in most cases. Our model's errors are almost one order or several orders of magnitude lower than theirs. When using the activation function ``softmax'' to calculate the validation test (the first column of the figure), we can see that the error of our model decreases to a stable state after only a few epochs, while Hutchinson's model needs much longer epochs to achieve the effect of our model. For the case of the activation function ``softsign'' and $S(0)/K = 0.5$ (the second column of the figure), our model has reached a local minimum error within 20 epochs. As the epoch increases, the error of our model has also increased because of overfitting. Even so, Hutchinson's model did not reach our minimum error within 100 epochs.

\subsection{Real market data}
We consider two real market datasets: the one from the SPX calls on Cboe in the USA, the other from the CSI 300 Index call options (IO-C) on CFFEX in China\footnote{We note that IO contracts are examples of Asian options. But data-driven models can ignore this fact. }. SPX contracts used are that expired on September 20th, 2019, October 18th, 2019, November 15th, 2019, December 20th, 2019, January 17th, 2020, February 21st, 2020, March 20th, 2020, and June 19th, 2020. We take all trade-based minute bars with positive volume from August 8th, 2019 to September 6th, 2019. The close price of each bar generates a sample point, without taking into account of multiplicities. Finally, there are 12323 examples in the dataset. In each trial, we randomly split them into 9858 for training and 2465 for validation.

Similarly, we use minute bars of IO2002, IO2003, IO2006, and IO2009 traded from January 2, 2020 to February 7, 2020. There are 63939 examples in the dataset. In each trial, the first $50\%$ of them are used for training model parameters, and the last $50\%$ are used for testing. We no longer randomly shuffle the data, and thus the distribution of the test set will be significantly different from the training set. That is why we call it the validation dataset in the SPX experiment and the test dataset in the IO experiment.

\begin{figure}[!t]
\centering
\includegraphics[width=3.5in]{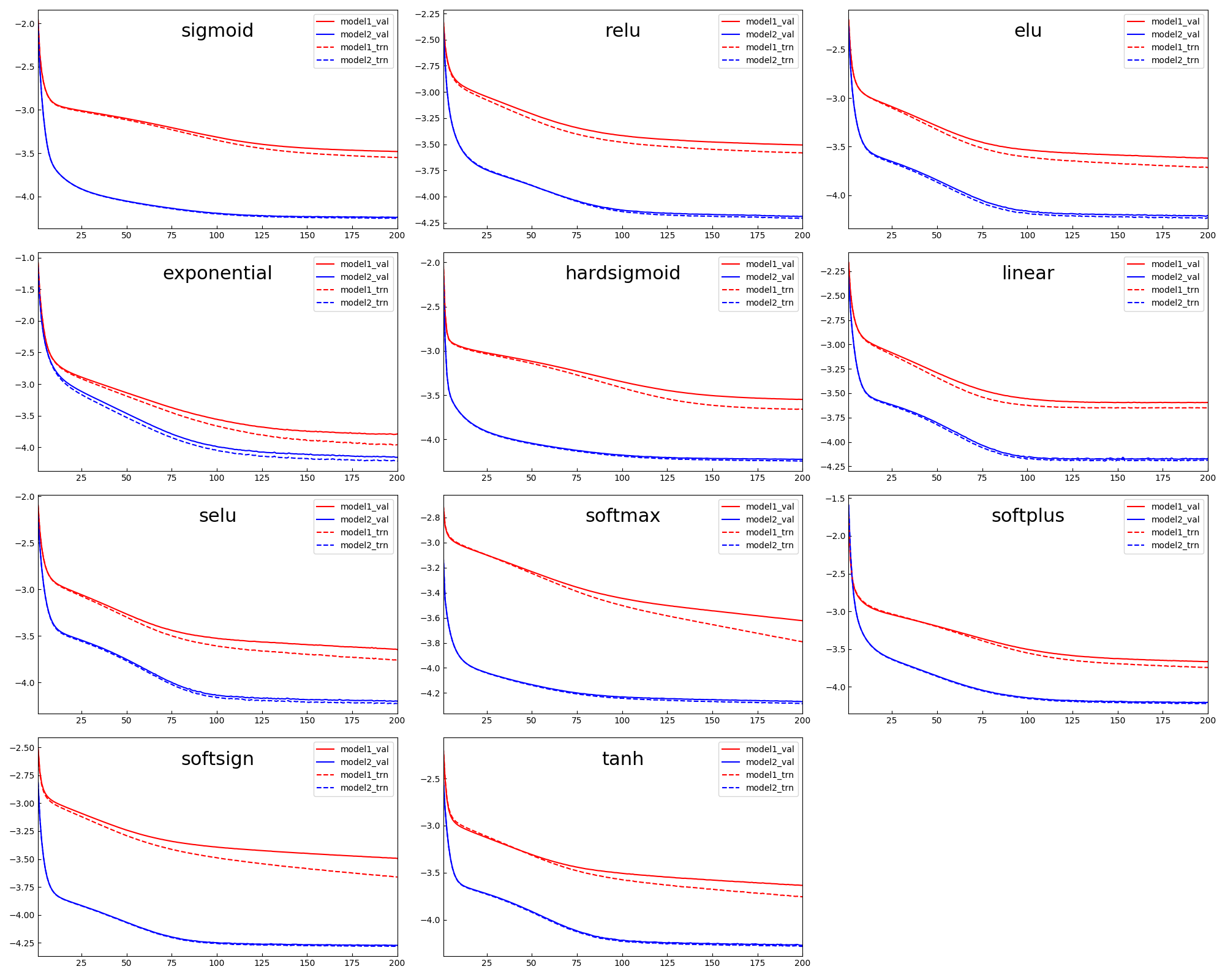}
\caption{The results of SPX Options. Each subfigure is the result of an activation function, with the function name written in the subfigure. The red lines represent the results of model (\ref{model1}) and the blue lines represent the results of model (\ref{model2}). The dashed lines represent the training set, and the solid lines represent the validation set. The x-axis value of each subfigure is epoch, and the y-axis value is $\log_{10}(\mathrm{MSE})$.}
\label{realresult}
\end{figure}

\begin{figure}[!t]
\centering
\includegraphics[width=3.5in]{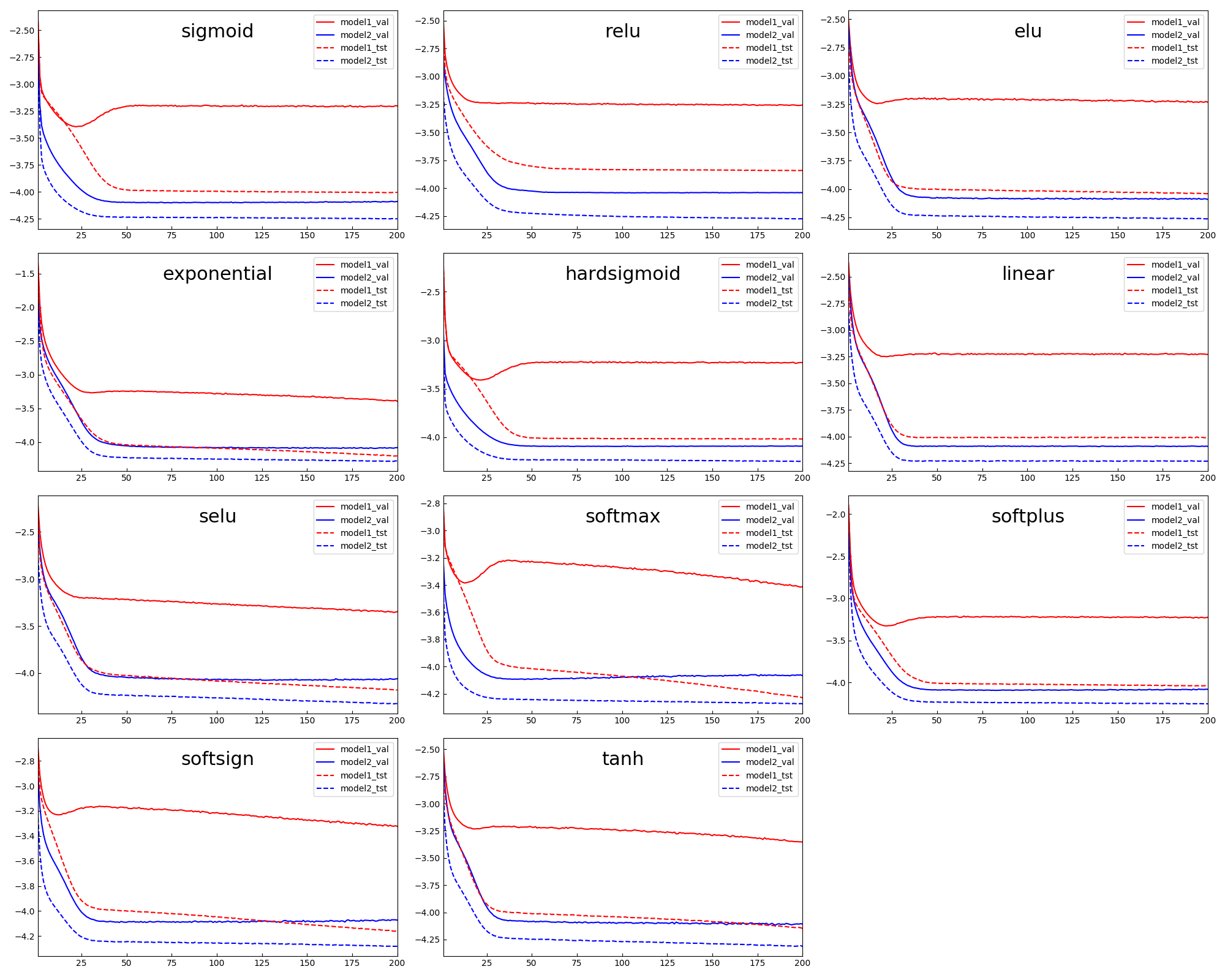}
\caption{The results of CSI 300 Index Options. Each subfigure is the result of an activation function, with the function name written in the subfigure. The red lines represent the results of model (\ref{model1}) and the blue lines represent the results of model (\ref{model2}). The dashed lines represent the training set, and the solid lines represent the test set. The x-axis value of each subfigure is epoch, and the y-axis value is $\log_{10}(\mathrm{MSE})$.}
\label{IOresult}
\vspace{-0.8cm}
\end{figure}

A total of 1000 trials are run for each activation function on each dataset to obtain the statistical average results. Fig. \ref{realresult} shows the numerical results of SPX options, and Fig. \ref{IOresult} IO. Comparing the experimental results of Model (\ref{model1}) and Model (\ref{model2}), we derive the following conclusions:

\begin{itemize}
\item Model (\ref{model2}) outperforms Model (\ref{model1}) in terms of convergence rate.

\item Model (\ref{model2}) outperforms Model (\ref{model1}) in terms of generalization performance.
\end{itemize}

The rest of this paper attempts to explain why our model is likely to deliver better generalization performance in experiments from the viewpoint of expressivity.

\section{Boundary behavior of the Black Scholes formula}\label{sectionbs}

A European stock option $C$ is a contract that allows the buyer to buy the underlying stock with a fixed strike price $K$ at the expiration date $T$. Let $\sigma, r, q$ be the volatility of the underlying, the risk-free rate of interest, and the dividend yield. Under assumptions of Black-Scholes model \cite{BS, Merton}, the price of the underlying $C(t)$ satisfies $C(t)/K = se^{-q\tau}N(d_1) - e^{-r\tau}N(d_2)$,
where $s \mathopen=\mathclose S(t)/K$, $d_1 \mathopen=\mathclose \frac{\log{s} + (r-q+\sigma^2/2) \tau}{\sigma \sqrt{\tau}}$,  $d_2 \mathopen=\mathclose \frac{\log{s} + (r-q-\sigma^2/2) \tau}{\sigma \sqrt{\tau}}$, and $
N(x) \mathopen=\mathclose \frac{1}{\sqrt{2\pi}} \int_{-\infty}^{x} e^{-t^2/2} dt$. With fixed $r, q$, and $\sigma$, the decision function $f$ of Model (\ref{model1}) is given by

\vspace{-1\baselineskip}
\begin{small}
\begin{align}\label{deff}
f(s, \tau) := se^{-q\tau}N(d_1) - e^{-r\tau}N(d_2),
\end{align}
\end{small}%
and the decision function $g$ of model (\ref{model2}) is given by

\vspace{-1\baselineskip}
\begin{small}
\begin{align}\label{defg}
g(s, \tau):= f(s, \tau) - (se^{-q \tau}-e^{-r \tau})^+.
\end{align}
\end{small}%
Both $f$ and $g$ are well defined on $\RR^+ \times [0,1]$. Although the above formula is stated for stock options, it also holds for the SPX index option. For the index can be regarded as a stock paying a dividend yield. Using the well-known trick $\left(\int_{t}^{+\infty} e^{-\frac{x^2}{2}} dx \right)^2
=  \int_{t}^{+\infty} e^{-\frac{x^2}{2}} dx \cdot \int_{t}^{+\infty} e^{-\frac{y^2}{2}} dy $, it is straightforward to verify
\begin{lemma}\label{lemma1}
If $t>0$ then $
\frac{1}{\sqrt{2 \pi}} \int_{t}^{+\infty} e^{-\frac{x^2}{2}} dx < \frac{1}{2}e^{-\frac{t^2}{2}}.$
\end{lemma}

%\begin{IEEEproof}
%It follows from
%\begin{align*}
%&\left( \frac{1}{\sqrt{2 \pi}} \int_{t}^{+\infty} e^{-\frac{x^2}{2}} dx \right)^2 \\
%&= \frac{1}{\sqrt{2 \pi}} \int_{t}^{+\infty} e^{-\frac{x^2}{2}} dx \cdot \frac{1}{\sqrt{2 \pi}} \int_{t}^{+\infty} e^{-\frac{y^2}{2}} dy \\
%&< \frac{1}{8 \pi}   \int_{x^2+y^2>2t^2} e^{-\frac{x^2+y^2}{2}} dx dy \\
%&=  \frac{1}{8 \pi} \int_0^{2\pi} \int_{\sqrt{2}t}^{+\infty} e^{-\frac{r^2}{2}} rdr d\theta \\
%&= \frac{1}{4} e^{-t^2}.
%\end{align*}
%\end{IEEEproof}

In the next proprosition, we prove that $g$ is integrable over $\RR^+ \times [0,1]$. For all $r\in (-\infty, \infty)$, $q \ge 0$, and $\sigma >0$, we have
\begin{proposition}\label{lemmalp}
For all $p \in [1, +\infty]$, $g \in L^p(\RR^+ \times [0,1])$.
\end{proposition}
\begin{IEEEproof}
For fixed $r, q, \sigma$, there exists $S^{'}>\max{(1, e^{q-r})}$ such that for all $0 \le \tau \le 1$ and $s > S^{'}$,

\vspace{-1\baselineskip}
\begin{small}
\begin{align*}
d_1 &\ge d_2 = \sigma^{-1}\tau^{-1/2}(\log{s} + (r-q-\sigma^2/2) \tau) \\
&> C_1 \log{s} + C_2 \\
&> C_3 \log{s},
\end{align*}
\end{small}%
where $C_1>0, C_2 \in (-\infty, +\infty)$, and $C_3>0$. The function $h$ defined by $h(\tau) := e^{(q-r) \tau}$ satisfies $h(0) =1$ and $h(1) = e^{q-r}$. If $s>S^{'}$, then $s \ge \max(h(0), h(1))$. Because $h$ is monotone, $s \ge h(\tau)$ for all $\tau \in [0,1]$, and thus $se^{-q\tau}-e^{-r\tau}\ge0$. By Lemma \ref{lemma1}, for all $s>S^{'}$ and $q \in [0,1]$,

\vspace{-1\baselineskip}
\begin{small}
\begin{align*}
|g(s, \tau)| &= |se^{-q\tau}N(d_1) - e^{-r\tau}N(d_2) -  (se^{-q \tau}-e^{-r \tau})^+| \\
&= |se^{-q\tau}(N(d_1) - 1) - e^{-r\tau}(N(d_2) - 1)| \\
&\leq  s(1-N(d_1)) + \max{\{1, e^{-r}\}}(1 - N(d_2)) \\
&< C_4 (se^{-d_1^2/2}+  e^{-d_2^2/2}) \\
&< 2C_4 s e^{- C_5 \log^2s} \\
&= 2C_4 s^{1- C_5 \log{s}},
\end{align*}
\end{small}%
where $C_4, C_5$ are positive constants. There exists $S^{''}>S^{'}$ such that for all $s > S^{''}$, $C_5 \log{s} > 3$, and therefore $|g(s, \tau)| < 2C_4 s^{-2}$, and consequently

\vspace{-1\baselineskip}
\begin{small}
\begin{align}\label{lp1}
g \cdot I_{[S^{''}, \infty) \times [0,1]} \in  L^p(\RR^+ \times [0,1]).
\end{align}
\end{small}%
On the other hand, for all $(s, \tau) \in (0, S^{''}) \times [0,1]$, we have

\vspace{-1\baselineskip}
\begin{small}
\begin{align*}
|g(s, \tau)| &= |se^{-q\tau}N(d_1) - e^{-r\tau}N(d_2) - (se^{-q \tau}- e^{-r \tau})^+| \\
&\leq  sN(d_1)+\max{(1, e^{-r})} N(d_2)+s+\max{(1, e^{-r})} \\
&\leq  2S^{''}+2\max{(1, e^{-r})}.
\end{align*}
\end{small}%
The above inequality implies that $g$ is bounded on $(0, S^{''}) \times [0,1]$, and thus

\vspace{-1\baselineskip}
\begin{small}
\begin{align}\label{lp2}
g \cdot I_{(0, S^{''}) \times [0,1]} \in  L^p(\RR^+ \times [0,1]).
\end{align}
\end{small}%
By (\ref{lp1}) and (\ref{lp2}), $g \in  L^p(\RR^+ \times [0,1]).$
\end{IEEEproof}

\section{Universal approximation theorem in $L^2(\RR \times [0,1])$}\label{sectionnn}
Let there be $y \in \RR^n$, $\theta \in \RR$, and a function $\psi$ defined on $\RR$. We set

\vspace{-1\baselineskip}
\begin{small}
\begin{align}\label{actions}
\psi^{\tau_{\theta}}:&= x \in \RR \mapsto \psi(x+\theta) \nonumber \\
\psi^{\delta_{y}}:&= x \in \RR^{n} \mapsto \psi(\langle y, x\rangle)  \nonumber \\
\psi^{\tau_{\theta} \delta_{y}} &= (\psi^{\tau_{\theta}})^{\delta_{y}},
\end{align}
\end{small}%
and let $\mathcal{S}_{n}(\psi)$ denote the space of the following linear combinations

\vspace{-1\baselineskip}
\begin{small}
\begin{align}\label{Sn}
\sum_{i=1}^{k}t_i\psi^{\tau_{\theta_i} \delta_{y_i}}, k \in \NN, y_i \in \RR^{n}, \theta_i \in \RR, t_i \in \RR.
\end{align}
\end{small}%
In the rest of the paper, $\phi$ is given by the logistic function $\phi(x) = 1/(1+e^{-x}).$
To prove the universal approximation theorem in the function space $L^2(\RR \times [0,1])$, we need the following estimate concerning the boundary behavior of $\phi^{\tau_{\theta_1} \delta_{y}}- \phi^{\tau_{\theta_2} \delta_{y}}$.
\begin{lemma}\label{boundarybehavior}
Let $\Omega = \RR \times [0,1]$ and $y = (y_1, y_2) \in \RR^2$.

\noindent 1, If $p \in [1, +\infty]$ and $\theta_1, \theta_2 \in \RR$, then

\vspace{-1\baselineskip}
\begin{small}
\begin{align*}
\phi^{\tau_{\theta_1} \delta_{y}}- \phi^{\tau_{\theta_2} \delta_{y}}\in  L^{p}(\Omega).
\end{align*}
\end{small}%
2, If $y_1 \neq 0$, then there exist positive constants $B, X, C$ such that for all $\beta > B, |x_1| > X, x=(x_1, x_2) \in \Omega$,

\vspace{-1\baselineskip}
\begin{small}
\begin{align*}
|(\phi^{\tau_{\beta \theta_1} \delta_{\beta y}}- \phi^{\tau_{\beta \theta_2} \delta_{\beta y}})(x)| \leq B C e^{-0.5 B |y_1 \cdot x_1|}.
\end{align*}
\end{small}%
\end{lemma}
\begin{IEEEproof}
Given $\beta>0$, by the mean value theorem, there exists $\theta_{\beta} \in [\theta_1, \theta_2]$ such that

\vspace{-1\baselineskip}
\begin{small}
\begin{align*}
&\phi^{\tau_{\beta \theta_1} \delta_{\beta y}}(x) -  \phi^{\tau_{\beta \theta_2} \delta_{\beta y}}(x) \\
=& \frac{\beta (\theta_1 - \theta_2) e^{- \beta \langle y, x \rangle -\beta \theta_{\beta}}}{(1+e^{- \beta \langle y, x \rangle -\beta \theta_{\beta}})^2} \\
=&  \frac{\beta (\theta_1 - \theta_2)}{(1+e^{- \beta y_1x_1 - \beta y_2 x_2-\beta \theta_{\beta}}) (1+e^{\beta y_1x_1 + \beta y_2 x_2+\beta \theta_{\beta}})}.
\end{align*}
\end{small}%
There exists a constant $C_6$ such that for all $x_2 \in [0,1]$ and $\theta \in [\theta_1, \theta_2]$,

\vspace{-1\baselineskip}
\begin{small}
\begin{align*}
|y_2 x_2 +\theta| < C_6.
\end{align*}
\end{small}%
Let $X = 2C_6/|y_1|$. Then for all $|x_1|>X$ and $\theta \in  [\theta_1, \theta_2]$,

\vspace{-1\baselineskip}
\begin{small}
\begin{align*}
|y_1 x_1 + (y_2 x_2  + \theta)| & \ge   |y_1 x_1| -  |y_2 x_2  + \theta| \\
&> |y_1 x_1| -  C_6 \\
&= |y_1 x_1| - 0.5 X |y_1| \\
&\ge |y_1 x_1| - 0.5 |y_1 x_1| \\
&= 0.5 |y_1 x_1|.
\end{align*}
\end{small}%
Therefore, for all $\beta > 0$,

\vspace{-1\baselineskip}
\begin{small}
\begin{align*}
& \ \ \ \ (1+e^{- \beta y_1 x_1 - \beta  y_2 x_2 -\beta \theta_{\beta}}) (1+e^{\beta y_1 x_1 + \beta y_2 x_2  +\beta \theta_{\beta}}) \\
&= (1+e^{\beta |y_1 x_1 + (y_2 x_2 + \theta_{\beta})|})  (1+e^{ -\beta |y_1 x_1 + (y_2 x_2 + \theta_{\beta})|}) \\
&\ge 1+e^{\beta |y_1 x_1+ (\langle y_2, x_2 \rangle +\theta_{\beta})|} \\
&> e^{0.5 \beta |y_1 \cdot x_1|}.
\end{align*}
\end{small}%
Consequently, there exists a constant $C_7>0$ such that for all $|x_1| > X$ and $\beta > 0$,

\vspace{-1\baselineskip}
\begin{small}
\begin{align}\label{1209}
|\phi^{\tau_{\beta \theta_1} \delta_{\beta y}}(x) -  \phi^{\tau_{\beta \theta_2} \delta_{\beta y}}(x)| \leq \beta C_7 e^{-0.5 \beta |y_1 x_1|}.
\end{align}
\end{small}%
The first part of our lemma is derived from (\ref{1209}) by assigning $\beta=1$. Let $B = 2/|y_1X|$.  For all $\beta > B$ and $|x_1|>X$, we have

\vspace{-1\baselineskip}
\begin{small}
\begin{align*}
&\ \ \ \ \partial(\beta C_7 e^{-0.5\beta|y_1 x_1|})/\partial \beta \\
&= C_7 e^{-0.5\beta|y_1 x_1|} -  0.5C_7 \beta |y_1 x_1|e^{-0.5\beta|y_1 x_1|} \\
&\leq (C_7 - C_7 |y_1| X/|y_1X|) e^{-0.5\beta|y_1 x_1|} \\
&= 0.
\end{align*}
\end{small}%
The above inequality implies that if  $\beta > B$ and $|x_1|>X$, then $\beta C_7 e^{-0.5\beta|y_1 x_1|}$ is monotonically decreasing in $\beta$. In particular, $\beta C_7 e^{-0.5 \beta |y_1 x_1|}\leq B C_7 e^{-0.5 B |y_1 x_1|}$.  Taking $C=C_7$, for all $\beta > B$ and $|x_1|>X$, we have

\vspace{-1\baselineskip}
\begin{small}
\begin{align*}
|\phi^{\tau_{\beta \theta_1} \delta_{\beta y}}(x) -  \phi^{\tau_{\beta \theta_2} \delta_{\beta y}}(x)| \leq B C e^{-0.5 B |y_1 x_1|},
\end{align*}
\end{small}%
which leads to the second part of the lemma.
\end{IEEEproof}

The Fourier transform $F \mapsto \widehat{F}$ maps $L^1(\RR^n)$ into $L^{\infty}(\RR^n)$, and it is an isometry of $L^2(\RR^n)$. Let there be $\theta \in S^1$ represented by $(\cos{\omega}, \sin{\omega})$. Its orthogonal complement is defined by $\theta^{\perp} = (-\sin{\omega}, \cos{\omega})$. The x-ray transform $P_{\theta}F$ of $F \in L^1(\RR^2)$ is a family of functions indexed by $\theta \in S^1$ and defined by: for $s \in \RR$,

\vspace{-1\baselineskip}
\begin{small}
\begin{align}\label{defxray}
P_{\theta}F(s\theta^{\perp}) = \int_{\RR}F(s\theta^{\perp}+t\theta)dt.
\end{align}
\end{small}%
The projection slice theorem (Theorem 2.2 in \cite{FARIDANI}) for the x-ray transform of $F$ tells us

\vspace{-1\baselineskip}
\begin{small}
\begin{align*}
\widehat{P_{\theta}F}(s\theta^{\perp}) = \widehat{F}(s\theta^{\perp}).
\end{align*}
\end{small}%
Let $\lambda_n$ denote the Lebesgue measure on $\RR^n$, and $\lambda_{S^1}$ the canonical measure on $S^1$ with volume $2\pi$. Using the Hahn-Banach theorem as in \cite{Cybenko} and based on some tedious estimates, we shall prove
\begin{theorem}\label{maintheorem}
Let $\Omega = \RR \times [0,1]$ and $\phi$ be the logistic activation function. Then,

\vspace{-1\baselineskip}
\begin{small}
\begin{align*}
\overline{\mathcal{S}_{2}(\phi) \cap L^{2}(\Omega)} = L^{2}(\Omega).
\end{align*}
\end{small}%
\end{theorem}
\begin{IEEEproof}
Suppose the theorem is not true, then by the Hahn-Banach Theorem, there exists a nonzero real valued function $h \in L^2(\Omega)$ such that for all  $G \in \mathcal{S}_{2}(\phi) \cap  L^{2}(\Omega)$,

\vspace{-1\baselineskip}
\begin{small}
\begin{align}\label{0402}
\int_{\Omega} G(x) h(x) dx = 0.
\end{align}
\end{small}%
By Lemma \ref{boundarybehavior}, for all $y = (y_1, y_2) \in \RR \times [0,1]$ that satisfies $y_1 \neq 0$, and $\rho \in \RR$, we have $\phi^{\tau_{0} \delta_{y}}- \phi^{\tau_{\rho} \delta_{y}} \in L^2(\Omega)$. By using (\ref{0402}),

\vspace{-1\baselineskip}
\begin{small}
\begin{align}\label{aaaa}
\int_{\Omega} (\phi^{\tau_{0} \delta_{y}}- \phi^{\tau_{\rho} \delta_{y}})(x) h(x) dx = 0.
\end{align}
\end{small}%
For $\rho \in \RR$, we define

\vspace{-1\baselineskip}
\begin{small}
\begin{align}\label{defgamma}
\gamma_{y, \rho}(x)=  \lim_{\beta \to +\infty} (\phi^{\tau_{0} \delta_{\beta y}}- \phi^{\tau_{\beta \rho} \delta_{\beta y}})(x).
\end{align}
\end{small}%
Let $B, X, C$ be constants that fulfill the second part of lemma \ref{boundarybehavior}.  Set $\Omega_{-} = (-X, X) \times [0,1]$, and $\Omega_{+} = \Omega \setminus \Omega_{-}$; and consider the following functions on $\Omega$,

\vspace{-1\baselineskip}
\begin{small}
\begin{align}\label{defRS}
R: x \in \Omega &\mapsto B C e^{-0.5 B |y_1x_1|}, \nonumber \\
S: x \in \Omega &\mapsto  R(x) |h(x)| I_{\Omega_{+}}(x) +  2|h(x)| I_{\Omega_{-}}(x),
\end{align}
\end{small}%
where $I$ is the indicator function. It is clear that $R$ is an $L^2$ integrable function, and $S$ is a non-negative function. Moreover,

\vspace{-1\baselineskip}
\begin{small}
\begin{align*}
& \int_{\Omega} S(x) dx = \int_{\Omega_{+}} R(x) |h(x)|dx + \int_{\Omega_{-}}2|h(x)|dx  \\
&\leq  \left(||R||_{L^2(\Omega_{+})} ||h||_{L^2(\Omega_{+})}\right)^{1/2} +  2 \left( \lambda_{2}(\Omega_{-})||h(x)||_{L^2(\Omega_{-})}\right)^{1/2},
\end{align*}
\end{small}%
and therefore $S \in L^1(\Omega)$. By Lemma \ref{boundarybehavior}, for $x \in \Omega_{+}$ and $\beta > B$, we have

\vspace{-1\baselineskip}
\begin{small}
\begin{align*}
|(\phi^{\tau_{0} \delta_{\beta y}}- \phi^{\tau_{\beta \rho} \delta_{\beta y}}) \cdot h(x)| < R(x)|h(x)|=S(x).
\end{align*}
\end{small}%
Because $\phi$ takes value in $(0, 1)$, for $x \in \Omega_{-}$ we also have

\vspace{-1\baselineskip}
\begin{small}
\begin{align*}
|(\phi^{\tau_{0} \delta_{\beta y}}- \phi^{\tau_{\beta \rho} \delta_{\beta y}}) \cdot h(x)| < 2|h(x)|=S(x).
\end{align*}
\end{small}%
Therefore for all $\beta>B$, the family of functions $(\phi^{\tau_{0} \delta_{\beta y}}- \phi^{\tau_{\beta \rho} \delta_{\beta y}}) \cdot h$ are uniformally bounded by  $S \in L^1(\Omega)$. The Lebesgue dominated convergence theorem, together with  (\ref{aaaa}), gives rise to the vanishing of the following integrals,

\vspace{-1\baselineskip}
\begin{small}
\begin{align*}
\int_{\Omega}\gamma_{y, \rho}(x) h(x) dx &= \int_{\Omega} \lim_{\beta \to \infty} (\phi^{\tau_{0} \delta_{\beta y}}- \phi^{\tau_{\beta \rho} \delta_{\beta y}}) h(x) dx \\
&=  \lim_{\beta \to \infty} \int_{\Omega} (\phi^{\tau_{0} \delta_{\beta y}}- \phi^{\tau_{\beta \rho} \delta_{\beta y}}) h(x) dx\\
 &= 0.
\end{align*}
\end{small}%

Furthermore, the function $\gamma_{y, \rho}$ satisfies  $\gamma_{y, \rho}(x)=  \overline{\gamma}_{\rho}(\langle y, x \rangle)$ for all $x \in \RR^{2}$, where $\overline{\gamma}_{\rho}= - I_{(-\rho, 0)} - 0.5 I_{\{-\rho\}} - 0.5I_{\{0\}}$ if $\rho>0$ and $\overline{\gamma}_{\rho}=  I_{(0, -\rho)} + 0.5 I_{\{-\rho\}} + 0.5I_{\{0\}}$ if $\rho < 0$. For all $y=(y_1, y_2)$ with $y_1 \neq 0$, and $\rho > 0$, we have

\vspace{-1\baselineskip}
\begin{small}
\begin{align*}
\int_{\Omega} I_{[-\rho, 0]}(\langle y, x \rangle) h(x) dx & =  - \int_{\Omega}\overline{\gamma}_{\rho}(\langle y, x \rangle) h(x) dx \\
& =  - \int_{\Omega}\gamma_{y, \rho}(x) h(x) dx = 0.
\end{align*}
\end{small}%
Similarly, for all $y=(y_1, y_2)$ with $y_1 \neq 0$, and $\rho > 0$, we have $\int_{\Omega} I_{[0, \rho]}(\langle y, x \rangle) h(x) dx = 0$. By setting $h|_{\RR^2 \setminus \Omega} = 0$, $h$ can be regarded as a function in $L^2(\RR^2)$. We thus have proved for all $y = (y_1, y_2)$ with $y_1 \neq 0$, and $\rho >0$,

\vspace{-1\baselineskip}
\begin{small}
\begin{align}\label{1210}
\int_{\RR^2} I_{[0, \rho]}(\langle y, x \rangle) h(x) dx = \int_{\RR^2} I_{[-\rho, 0]}(\langle y, x \rangle) h(x) dx = 0.
\end{align}
\end{small}%
Given $\theta =  (\cos{\omega}, \sin{\omega}) \in S^1  \setminus \{ (\pm 1, 0) \}$, where $\omega \in (0,\pi) \cup (\pi, 2\pi)$, we consider the transformation given by $x =\varphi(s, t)=t\theta^{\perp}+s\theta$, i.e.,

\vspace{-1\baselineskip}
\begin{small}
\begin{align}\label{changeofvariable}
\varphi(s, t) = (s\cos{\omega}-t\sin{\omega}, s\sin{\omega}+t\cos{\omega}).
\end{align}
\end{small}%
It is a one-one map between $\{ (x_1, x_2) \in \RR^2 \}$ and $\{ (s, t) \in \RR^2 \}$.  Use this change of variable and take $y=\theta^{\perp}$, (\ref{1210}) gives rise to

\vspace{-1\baselineskip}
\begin{small}
\begin{align*}
0  &=  \int_{\RR^2} I_{[0, \rho]}(\langle \theta^{\perp}, t\theta^{\perp}+s\theta \rangle) h(t\theta^{\perp}+s\theta) dsdt  \\ &= \int_{\RR^2} I_{[0, \rho]}(t) h(t\theta^{\perp}+s\theta) dsdt  \\
&= \int^{\rho}_{0} \int_{\RR} h(t\theta^{\perp}+s\theta) dsdt.
\end{align*}
\end{small}%
Taking the derivative of both sides of the above identity with respect to $\rho$, we have, for almost every $\rho>0$,

\vspace{-1\baselineskip}
\begin{small}
\begin{align}\label{importantpre}
\int_{\RR} h(\rho\theta^{\perp}+s\theta) ds = 0.
\end{align}
\end{small}%
By similar arguments, (\ref{importantpre}) also holds for almost every $\rho <0.$

\begin{figure}[!t]
\centering
\includegraphics[width=3.5in]{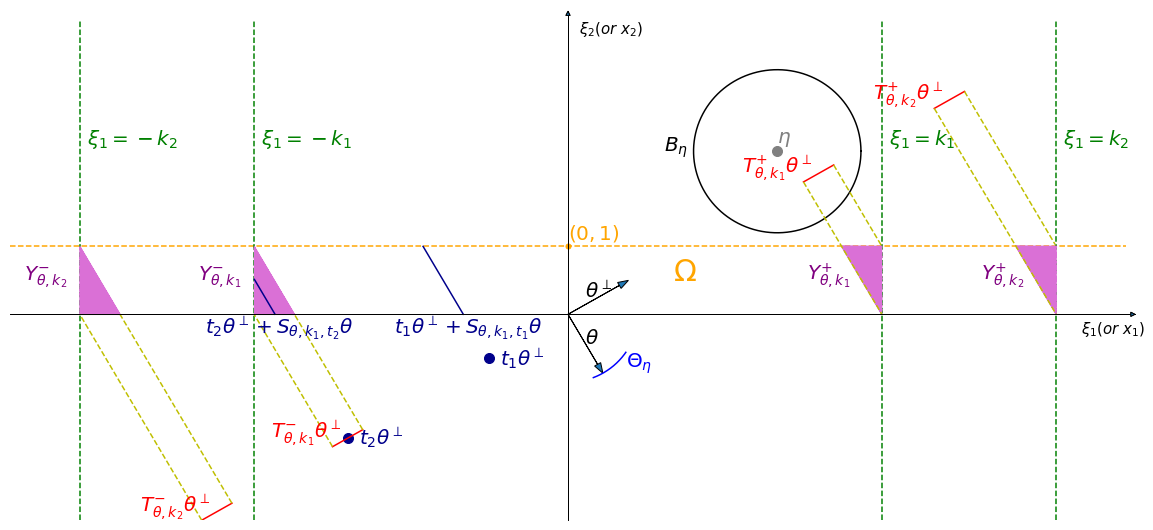}
\caption{Some Notations and ideas of the proof. We want to prove that $\widehat{h}(\eta)$ is zero. This follows from estimates of $\widehat{h_k}(\eta)$ and a limit argument. We use the fact that under our assumption  the value of $\widehat{h_k}$ on a small neighborhood of $\eta$ is only contributed by the value of $h$ on the set of small variations of $Y^{\pm}_{\theta, k}$ (the orchid area in the figure).}
\label{proofpng}
\vspace{-1cm}
\end{figure}

Before proceeding, we would like to mention that Fig. \ref{proofpng} might help the reader better understand the following arguments. For $k \in \NN$, set $\Omega_k = (-k, k) \times [0,1]$ and $h_k: = I_{\Omega_k} h$. Then $h_k$ is a function in $L^1(\RR^{2}) \cap L^2(\RR^{2})$. For $\xi \in \RR^2 \setminus \{0, 0\}$, let $\theta_{\xi}$ denote the unique point in $S^1$ satisfying $\xi \in \RR^{+} \theta_{\xi}^{\perp}$. For $\eta = (\eta_1, \eta_2) \in \RR^2$ with $\eta_1 > 0, \eta_2>0$, we set $B_{\eta} = \{\xi \in \RR^2: |\eta-\xi| \leq \max{\{\eta_1, \eta_2\}}/2\}$ and $\Theta_\eta = \{\theta_{\xi}: \xi \in B_{\eta} \}.$ There is a constant $C_{\eta}$ such that $|\xi| \leq C_{\eta}$ for all $\xi \in B_{\eta}.$ Moreover, $\Theta_{\eta}$, the blue arc in Figure \ref{proofpng}, is a compact subset of $ \{(\cos{\omega}, \sin{\omega}): \omega \in (3\pi/2, 2\pi)\} \subset S^1$. For any $\theta = (\cos{\omega}, \sin{\omega}) \in \Theta_{\eta}$, we set

\vspace{-1\baselineskip}
\begin{small}
\begin{align}\label{defalot}
T^{\pm}_{\theta, k} &= \{t \in \RR: (t \theta^{\perp}+\RR \theta) \cap (\{\pm k\} \times [0,1]) \neq \emptyset\} \nonumber\\
T^{\dag}_{\theta, k}&= T^{+}_{\theta, k} \cup T^{-}_{\theta, k} \nonumber\\
S_{\theta, k, t} &= \{s \in \RR: t \theta^{\perp}+s \theta \in \Omega_k\} \nonumber\\
Y^{\pm}_{\theta, k} &=   \{ \varphi(s, t): t \in T^{\pm}_{\theta, k}, s \in S_{\theta, k, t} \} \nonumber\\
Y^{\dag}_{\theta, k}&= Y^{+}_{\theta, k} \cup Y^{-}_{\theta, k},
\end{align}
\end{small}%
where $\varphi$ is defined in (\ref{changeofvariable}). Let $\theta = (\cos{\omega_{\theta}}, \sin{\omega_{\theta}})$ and $\lambda_1$ be the Lebesgue measure on $\RR^1$, then $\lambda_1(S_{\theta, k, t} ) \leq 1/|\sin{\omega_{\theta}}|$ (see Fig. \ref{proofpng}, $\lambda_1(S_{\theta, k, t})$ is less than or equal to the length of the segment $t_1\theta^{\perp}+S_{\theta, k_1,t_1}\theta$ there). Because $\Theta_{\eta}$ is  compact, $C^{'}_{\eta}:=\max{\{1/|\sin{\omega_{\theta}}|: \theta \in \Theta_{\eta} \}}$ is finite.  If $\theta \in \Theta_{\eta}, t \in \RR, k \in \NN$, then

\vspace{-1\baselineskip}
\begin{small}
\begin{align}\label{boundness}
\lambda_1(S_{\theta, k, t})<C^{'}_{\eta}.
\end{align}
\end{small}%
We claim that for almost every such $t \notin  T^{\dag}_{\theta, k}$,

\vspace{-1\baselineskip}
\begin{small}
\begin{align}\label{important}
\int_{\RR}h_k(t \theta^{\perp}+s\theta)ds  = 0.
\end{align}
\end{small}%
Indeed, if $S_{\theta, k, t} = \emptyset$ (which means the line $t\theta^{\perp}+\RR\theta$ is disjoint from $\Omega_k$), then $h_k(t \theta^{\perp}+\RR\theta)$ is constant zero and thus (\ref{important}) is true. Otherwise, $S_{\theta, k, t} \neq \emptyset$. Because $t \notin  T^{\dag}_{\theta, k}$, $(t\theta^{\perp}+\RR\theta) \cap \Omega_k = (t\theta^{\perp}+\RR\theta) \cap \Omega$). Therefore, $h_k|_{t \theta^{\perp}+\RR \theta} = h|_{t \theta^{\perp}+\RR \theta}$, and thus (\ref{important}) follows from (\ref{importantpre}).

By definition, (\ref{changeofvariable}) is one-one between $\{(s, t): t \in T^{\pm}_{\theta, k}, s \in S_{\theta, k, t} \}$ and $Y^{\pm}_{\theta, k_1}$, and also one-one between $\{(s, t): t \in T^{\dag}_{\theta, k}, s \in S_{\theta, k, t} \}$ and $Y^{\dag}_{\theta, k_1}$. It is clear (see Fig. \ref{proofpng}) that for positive integers $(k_2, k_1)$ that are sufficiently large,

\vspace{-1\baselineskip}
\begin{small}
\begin{align*}
Y^{+}_{\theta, k_2} = Y^{+}_{\theta, k_1} + (k_2-k_1, 0), \\
Y^{-}_{\theta, k_2} = Y^{-}_{\theta, k_1} - (k_2-k_1, 0).
\end{align*}
\end{small}%
Consequently for sufficiently large $k_2$ and $k_1$,

\vspace{-1\baselineskip}
\begin{small}
\begin{align}\label{shift}
\small{\bigcup}_{\theta \in \Theta_{\eta}}Y^{\pm}_{\theta, k_2} = \small{\bigcup}_{\theta \in \Theta_{\eta}}Y^{\pm}_{\theta, k_1} \pm (k_2-k_1, 0).
\end{align}
\end{small}%
As $\Theta_{\eta}$ is  compact, $\small{\bigcup}_{\theta \in \Theta_{\eta}}Y^{\pm}_{\theta, k}$ is also compact.  By (\ref{shift}), $\small{\bigcup}_{\theta \in \Theta_{\eta}}Y^{\pm}_{\theta, k}$ is shifted to infinity as $k$ goes to infinity. This implies $\lim\limits_{k \to \infty}||h||_{L^2(\cup_{\theta \in \Theta_{\eta}}Y^{\dag}_{\theta, k})} = 0.$ For any $\epsilon>0$, there exists $K$ such that if $k>K$ then

\vspace{-1\baselineskip}
\begin{small}
\begin{align}\label{gotoinfity}
\int_{\cup_{\theta \in \Theta_{\eta}}Y^{\dag}_{\theta, k}} h^2_k(x)dx < (2\pi C_{\eta}C^{'}_{\eta}\lambda_{S^1}(\Theta_{\eta}))^{-1}\epsilon.
\end{align}
\end{small}%
If $k>K$, by the projection slice theorem (\ref{boundness}), change of variable $\xi = t\theta^{\perp}$ (precisely $(\xi_1, \xi_2) = (-t\sin{\omega}, t\cos{\omega})$ which is one to one between $\RR^2 \setminus \{(0, 0)\}$ and $\RR^+ \times S^1$),

\vspace{-1\baselineskip}
\begin{small}
\begin{align*}
\int_{B_{\eta}} \left|\widehat{h_k}(\xi)\right|^2 d \xi &= \int_{B_{\eta}} \left|\widehat{P_{\theta_{\xi}}h_k}(\xi)\right|^2 d \xi \\
&= \int_{\Theta_{\eta}} \int_{ \{t>0: t\theta^{\perp} \in B_{\eta} \} } \left|\widehat{P_{\theta}h_k}(t \theta^{\perp})\right|^2t dt d \omega \\
&\leq C_{\eta} \int_{\Theta_{\eta}} \int_{ \{t>0: t\theta^{\perp} \in B_{\eta} \} }\left|\widehat{P_{\theta}h_k}(t \theta^{\perp})\right|^2 dt d \omega \\
&\leq C_{\eta} \int_{\Theta_{\eta}} \int_{\RR} \left|\widehat{P_{\theta}h_k}(t \theta^{\perp})\right|^2 dt d \omega.
\end{align*}
\end{small}%
By the fact that Fourier transform is an isometry of $L^2$ and the fact that $P_{\theta}h_k$ is real valued,

\vspace{-1\baselineskip}
\begin{small}
\begin{align*}
\int_{B_{\eta}} \left|\widehat{h_k}(\xi)\right|^2 d \xi &\leq C_{\eta} \int_{\Theta_{\eta}} \int_{\RR}(P_{\theta}h_k)^2(t \theta^{\perp}) dt d \omega  \\
&= C_{\eta} \int_{\Theta_{\eta}} \int_{\RR}  \left(\int_{\RR}h_k(t \theta^{\perp}+s\theta)ds \right)^2 dt d \omega.
\end{align*}
\end{small}%
By (\ref{important}), the Cauchy-Schwarz inequality, and (\ref{boundness}), we have

\vspace{-1\baselineskip}
\begin{small}
\begin{align*}
\int_{B_{\eta}} \left|\widehat{h_k}(\xi)\right|^2 d \xi &\leq C_{\eta} \int_{\Theta_{\eta}} \int_{T^{\dag}_{\theta, k}}  \left(\int_{\RR}h_k(t \theta^{\perp}+s\theta)ds \right)^2 dt d \omega \\
&=C_{\eta} \int_{\Theta_{\eta}} \int_{T^{\dag}_{\theta, k}}\left(\int_{S_{\theta, k, t}}h_k(t \theta^{\perp}+s\theta)ds\right)^2 dt d \omega \\
&\leq C_{\eta} C^{'}_{\eta}  \int_{\Theta_{\eta}} \int_{T^{\dag}_{\theta, k}}\int_{S_{\theta, k, t}}h_k^2(t \theta^{\perp}+s\theta) ds dt d \omega.
\end{align*}
\end{small}%
By the change of variable (\ref{changeofvariable}) and (\ref{gotoinfity}) we continue with

\vspace{-1\baselineskip}
\begin{small}
\begin{align*}
\int_{B_{\eta}} \left|\widehat{h_k}(\xi)\right|^2 d \xi &\leq C_{\eta} C^{'}_{\eta}  \int_{\Theta_{\eta}} \int_{Y^{\dag}_{\theta, k}} h_k^2(x) dx d \omega \\
&\leq   C_{\eta} C^{'}_{\eta}  \int_{\Theta_{\eta}} \int_{\cup_{\theta \in \Theta_{\eta}}Y^{\dag}_{\theta, k}} h_k^2(x) dx d \omega. \\
&\leq \epsilon.
\end{align*}
\end{small}%
This implies $ \lim\limits_{k \to \infty} ||\widehat{h_k}||^2_{L^2(B_{\eta})} =0$ and therefore $||\widehat{h}||^2_{L^2(B_{\eta})} =0$ for all $\eta \in \RR^+ \times \RR^+$. Similar arguments work for  $\eta \in \RR^+ \times \RR^-,\eta \in \RR^- \times \RR^+$ and $\eta \in \RR^- \times \RR^-$. Consequently $||\widehat{h}||^2_{L^2(\RR^2)} =0$, which contradicts to the fact that $h \neq 0$ in $L^2(\RR^2)$.
\end{IEEEproof}

The next corollary justifies the benefit of time value in the design of option neural network models. Let $f$ and $g$ be defined as in (\ref{deff}) and (\ref{defg}). We have
\begin{corollary}
Regarding $f$  and $g$ as functions on $\Omega = \RR \times [0,1]$, then $g \in \overline{\mathcal{S}_{2}(\phi) \cap L^{2}(\Omega)}$. Moreover, there exists no $\mathcal{N} \in \mathcal{S}_{2}(\phi)$ such that $||\mathcal{N}-f||_{L^2(\Omega)}$ is finite.
\end{corollary}
\begin{IEEEproof}
By Proposition \ref{lemmalp}, $g \in L^{2}(\Omega)$. By the above theorem, $g \in \overline{\mathcal{S}_{2}(\phi) \cap L^{2}(\Omega)}$.

If the second part of our corollary is not true, then there exists $\mathcal{N}_1 \in \mathcal{S}_{2}(\phi)$ such that $||\mathcal{N}_1-f||_{L^2(\Omega)}$ is finite. By the first part of our corollary, there exists $\mathcal{N}_2 \in \mathcal{S}_{2}(\phi)$ such that $||\mathcal{N}_2-g||_{L^2(\Omega)}$ is finite. Let $\mathcal{N}=\mathcal{N}_1-\mathcal{N}_2$ and $IV(s, \tau) := f(s, \tau) - g(s, \tau) = (se^{-q\tau} - e^{-r\tau})^+$, then $\mathcal{N} \in  \mathcal{S}_{2}(\phi)$ and $||\mathcal{N}-IV||_{L^2(\Omega)}$ is finite. This is impossible as $\mathcal{N}$ is a bounded function (our activation function is the sigmoid) and $IV(s, \tau)$ is asymptotically linear to $se^{-q\tau}$ for $s$ sufficiently large.
\end{IEEEproof}

This corollary implies that $g$ (but not $f$) can be approximated by superpositions of logistic function in $L^2(\RR^+ \times [0,1])$. A network that can not approximate the decision function on its full domain of definition is not supposed to generalize at tail events. In contrast, a network that can approximate the decision function on its full domain of definition has a better chance. Using the expressivity theory of neural networks, we provide a theoretical justification of the better generalization performance of Model (\ref{model2}).

\section{Conclusion}

%We are convinced that the dummy variable ``time value or option price'' should be a natural hyperparameter in the design of option network models.

This paper proposed a decision function based on time values to approach an old problem of option pricing. We would like to highlight the following contributions:

\begin{itemize}
\item We construct an explicit time value function for European calls, and prove that the time value normalized by the strike is an integrable function.
\item We first proposed using time value as the decision function for option pricing networks.
\item In experiments, our model is faster at learning and delivers better generalization performance.
\item We prove that a time value related function is better approximated by networks on an unbounded domain.
\end{itemize}

In particular, this paper discusses the last point in full details. Our work on the universal approximation theorem, together with our data experiments of an option model, suggests the following link between expressivity and generalization of neural networks: A learning framework with better approximation capability is likely to deliver better generalization performance.

\end{document}